\documentclass[namedreferences]{solarphysics}

\usepackage[hyperref,optionalrh]{spr-sola-addons} 
\usepackage{graphicx}        
\usepackage{color}           
\usepackage{breakurl}        

\newcommand{\arcsec}{$^{\prime\prime}$}

\begin{document}

\begin{article}
\begin{opening}

\title{Strong Blue Asymmetry in H$\alpha$ line as a Preflare Activity}

\author[addressref={aff1},email={chokh@astro.snu.ac.kr}]{Kyuhyoun Cho}
\author[addressref={aff1,aff2},corref,email={leej@astro.snu.ac.kr}]{Jeongwoo Lee}
\author[addressref={aff1}]{Jongchul Chae}
\author[addressref={aff2,aff3}]{Haimin Wang}
\author[addressref={aff3}]{Kwangsu Ahn}
\author[addressref={aff1}]{Heesu Yang}
\author[addressref={aff4}]{Eun-kyung Lim}
\author[addressref={aff1}]{Ram Ajor Maurya}

\address[id=aff1]{Astronomy Program, Department of Physics and Astronomy,
    				 Seoul National University, Seoul 08826, Korea}
\address[id=aff2]{Space Weather Research Laboratory,
             New Jersey Institute of Technology, University Heights,
             Newark, NJ 07102-1982, USA}
\address[id=aff3]{Big Bear Solar Observatory, Big Bear City,
             CA 92314-9672, USA}
\address[id=aff4]{Korea Astronomy and Space Science Institute,
         		Daejeon 305-348, Korea}

\runningauthor{Cho et al.}
\runningtitle{Blueshifted H$\alpha$ as a Preflare Activity}

\begin{abstract}
  Chromospheric activities prior to solar flares provide important clues to solar flare initiation, but are as yet poorly understood. We report a significant and rapid H$\alpha$ line broadening before the solar flare SOL2011-09-29T18:08, that was detected using the unprecedented high-resolution H$\alpha$ imaging spectroscopy with the {\it Fast Imaging Solar Spectrograph} (FISS) installed on the 1.6 m {\it New Solar Telescope} (NST) at Big Bear Solar Observatory.
The strong H$\alpha$ broadening extends as a blue excursion up to $-$4.5 \AA\ and as a red excursion up to 2.0 \AA\, which implies a mixture of velocities in the range of $-130$ km s$^{-1}$ to 38 km s$^{-1}$ by applying the cloud model, comparable to the highest chromospheric motions reported before. The H$\alpha$ blueshifted broadening lasts for about 4 minutes, and is temporally and spatially correlated with the start of a rising filament, which is later associated with the main phase of the flare as detected by the {\it Atmosphere Imaging Assembly} (AIA) onboard {\it Solar Dynamics Observatory} (SDO). The potential importance of this H$\alpha$ blueshifted broadening as a preflare chromospheric activity is briefly discussed within the context of the two-step eruption model.
\end{abstract}

\keywords{Flares, Pre-Flare Phenomena; Heating, in Flares; Spectrum, Visible}
\end{opening}

\section{Introduction}

The preflare phase refers to the early stage of the flare development process, generally prior to the impulsive hard X-ray emission \citep{Benz2002}. Since \citet{Bumba1959}'s first report, the preflare phase has been regarded as a key subject to understand the triggering mechanism of flares. The preflare phase is observed as local transient brightenings. A preflare brightening is smaller than a flare kernel, usually being a few dozen of arcseconds ({\it e.g.}, \citealt{Joshi2011}). It is mostly located in the vicinity of emerging flux regions \citep{Sterling2007}, eruptive filaments \citep{Chifor2007}, or canceling magnetic features \citep{Moon2004, Sterling2011}, which are roughly co-spatial with flare kernels that occur later. The typical time scale of a preflare brightening is a few dozen of minutes, being shorter than the main flare duration ({\it e.g.}, \citealt{Sterling2011}). These spatial and temporal characteristics of the brightenings were interpreted as heating due to localized and short-lived magnetic reconnection \citep{Joshi2013}.

Chromospheric motions can also be an important element for defining the preflare activity. There have been many studies reporting rising motions ({\it e.g.}, \citealt{Kundu1985, Sterling2011}) and oscillatory motions ({\it e.g.}, \citealt{Malville1981, Bocchialini2011}) within filaments before the flare occurrence. However, only a few spectroscopic studies on the preflare chromospheric motions have been reported. \citet{Canfield1990} observed a H$\alpha$ blueshifted absorption feature which was clearly assocated with an erupting preflare filament. \citet{Alikaeva2004} found a chromospheric upflow with speed higher than 10 km s$^{-1}$ in their preflare H$\alpha$ spectra obtained with the ATsU-26 solar telescope. They also found two crossing H$\alpha$ loops interact before the flare, after which the redistribution of velocities occurred in both the chromosphere and photosphere. \citet{Cauzzi1996} reported an upward motion a few seconds before the impulsive phase of the flare determined from lines originating in high chromospheric layers (Ca {\sc ii} K and H$\delta$) and from metallic lines (Si {\sc i} 3905, Fe {\sc i} multiplets 4 and 5). These motions together with a simultaneous strong emission were interpreted as implying early chromospheric heating prior to flares. Most recently, \citet{Leiko2015} studied a microflare using the H$\alpha$ spectrograph of the solar telescope, {\it T\'elescope H\'eliographique pour l'Etude du Magn\'etisme et des Instabilit\'es Solaires} (THEMIS), to find strong temporal variations of the line-of-sight (LOS) velocity in the chromosphere. The H$\alpha$ Doppler velocity changed from $-25$ km s$^{-1}$ at about 13 min before the microflare maximum to 5 km s$^{-1}$ at the maximum. The result suggests that there could be considerable mixing of both the magnitude and direction of the chromospheric motions.

There are also other types of chromospheric motions not limited to the preflare activity. Most widely studied are the chromospheric motions during the main phase of flares (see \citet{Falchi2006} and references therein). \citet{Svestka1962}, \citet{Ichimoto1984}, and others have revealed the red asymmetry of the H$\alpha$ line profiles in the initial phase of flares interpreted as chromospheric downflows with velocities of tens of km s$^{-1}$. \citet{Keys2011} measured LOS Doppler velocities of a C-class flare using the {\it Solar Optical Universal Polarimeter} (SOUP) that provides two-dimensional spectral information across the H$\alpha$ line profile at 7 wavelength positions. The first frame exhibited a red-shifted velocity of 6 km s$^{-1}$, and the subsequent frame showed a blue-shifted velocity of $-15$ km s$^{-1}$ at the same position.
\citet{Voort2009} studied the H$\alpha$ spectrogram of the quiet Sun to find a phenomenon called rapid blue excursions (RBE) which appears in absorption at a wide range of wavelengths from the line center to the far blue wing, typically extending up to $-2$ \AA~or $-90$ km s$^{-1}$.
These results suggest that LOS velocity appears in various settings and strong Doppler blueshifts indicative of violent upflows are observed as frequently as the redshifts indicating down-streaming of material from the corona.

In this paper we present spectroscopic observation of preflare activities in the chromosphere preceding the solar flare SOL2011-09-29T18:08, which shows a very strong blueshifted component in H$\alpha$ spectra. We have to note that the GOES data has a gap between 18:01 UT and 18:08 UT in which the flare maximum occurred, and the solar flare identifier refers to the time of resuming GOES measurement in the declining phase. Its soft X-ray class, C5.6, should therefore be regarded as an underestimation of its true strength.
The spectroscopic imaging observation was made using the {\it Fast Imaging Solar Spectrograph} (FISS) of the 1.6 meter {\it New Solar Telescope} (NST) at Big Bear Solar Observatory and the {\it Atmosphere Imaging Assembly} (AIA) onboard {\it Solar Dynamics Observatory} (SDO). This combination of instruments allows us to investigate not only the spectral properties but also spatial and temporal properties of the chromospheric motions with higher precision.

\section{Observations}

The FISS produced four-dimensional (spectral, temporal and two spatial dimensions) H$\alpha$ and Ca {\sc ii} 8542 \AA~data simultaneously in every scan. The details of the FISS instrument and  data processing were described by \citet{Chae2013b}. The FISS data that we took have the field of view (FOV) of 40\arcsec $\times$ 60\arcsec\ and the cadence of 66 seconds.
The spectral profile for the H$\alpha$ line is measured at 512 spectral positions and the spectral sampling is 0.019 \AA, and for the Ca {\sc ii} line, 502 spectral positions are used and the spectral sampling is 0.026 \AA. The wavelength calibration is based on the two lines H {\sc i} 6562.817 \AA~and Ti {\sc ii} 6559.580 \AA in the quiet region average profiles. We have used FISS observations from 17:12 UT to 18:40 UT, fully covering the flare including the preflare phase.

We also analyzed the data taken by the AIA \citep{Lemen2012} and the \textit{Helioseismic and Magnetic Imager} (HMI, \citealt{Schou2012}) onboard SDO. The time sequence of the 304 \AA~images from the SDO/AIA and of the LOS vector magnetogram from the SDO/HMI were used for checking the chromospheric and the photospheric features and their dynamics. We employed a time-distance map to measure the transverse motion and brightenings. We constructed the time-distance map of the SDO/AIA 304 \AA~images along the line that is oriented in the direction clearly displaying the initial movement of the chromospheric feature. The SDO/AIA 171 \AA~images were also used to investigate the configuration and the temporal variation of the coronal magnetic field lines.

We have checked hard X-ray data to confirm that the {\it Reuven Ramaty High Energy Solar Spectroscopic Imager} (RHESSI) data are missing for this time period. The Fermi Gamma-ray Burst Monitor (GBM) data are available and show the impulsive X-ray count flux from the solar flare up to 100 keV occurring around 18:03 UT. The GOES flare strength (C5.6) is certainly underestimated due to the missing GOES flux during the maximum phase. Small GBM peaks are found from 17:50 UT to 17:54 UT, which are not considered solar signals (not shown here). Presumably, they occurred at the time of a charged particle peak and another group of peaks with a similar shape re-appears at 18:32 UT possibly in association with the Fermi orbit (Brian Dennis 2015, private communication).

\section{Spatial Structure}

\begin{figure*}
\centering
  \includegraphics[width=1 \textwidth]{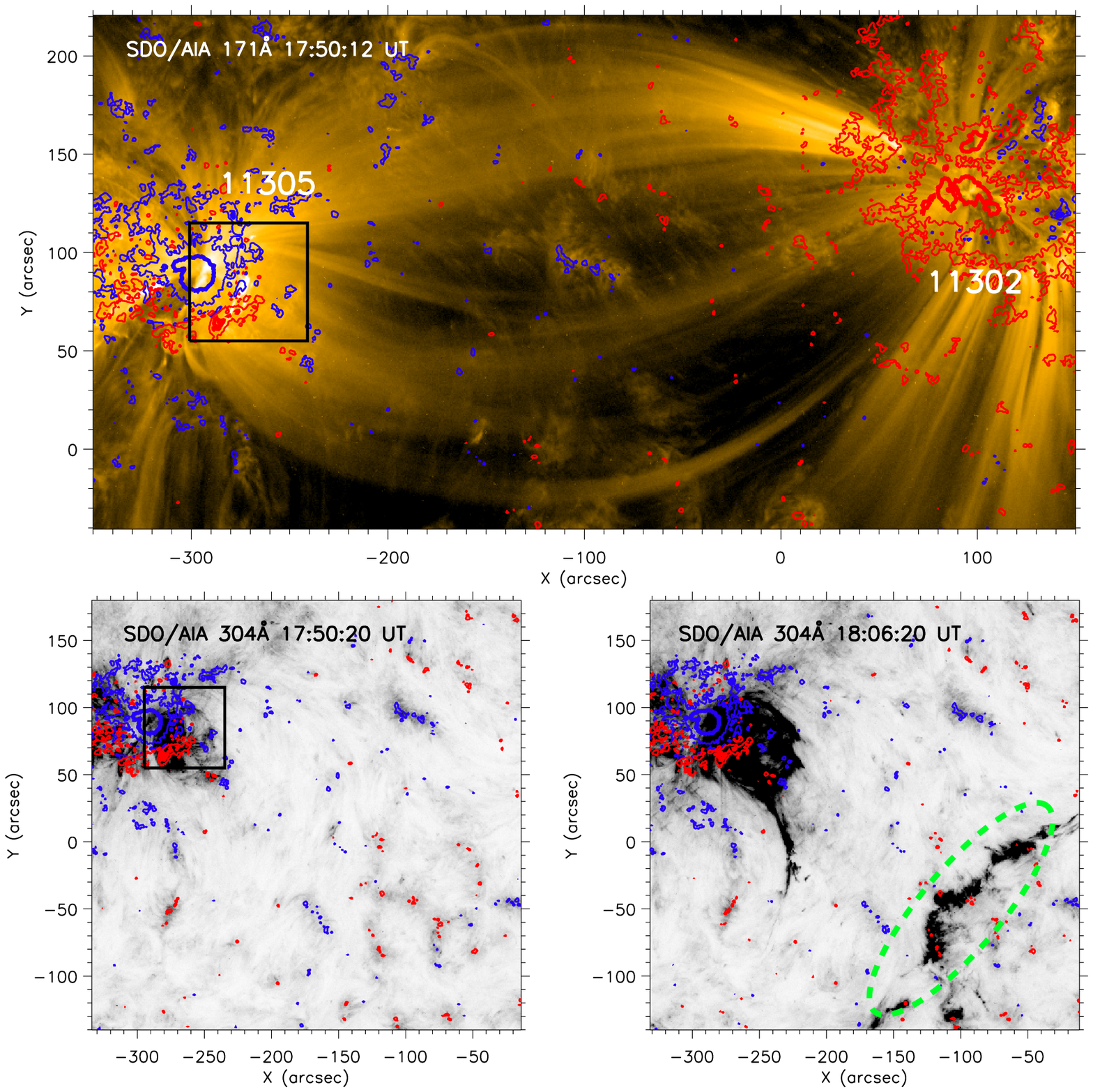}
         \caption{Top: Large FOV images of the source active region, NOAA 11305. Top: SDO/AIA 171 \AA\ image before the flare. Red (blue) contours represent positive (negative) polarity magnetic field, and the contour levels are in $\pm$100, $\pm$1000 G, respectively, from the SDO/HMI magnetogram. The same contour levels are used for magnetic fields in all other figures. The black box indicates the FOV of the panels displayed  in Figure 2. Bottom: SDO/AIA 304 \AA~negative images before (left) and after (right) the main flare occurrence. The dashed green ellipse represents the remote flare kernels.
              }
         \label{large fov}
   \end{figure*}

Figure \ref{large fov} shows that the observed active region is the trailing part of a larger-scale bipolar structure. The active region was connected to another active region (NOAA 11302) located about 300 Mm east of it by high-lying coronal loops clearly seen in SDO/AIA 171 \AA~images. The main polarity of the active region is negative, and that of its leading partner is positive. The flare caused disturbances in some of these coronal loops, but did not permanently disrupt them. There was no report of a coronal mass ejection (CME) associated with this flare. From the SDO/AIA 304 \AA~images, we also found the remote flare kernels appeared 150 Mm southeast of the active region when the flare occurred. Even though the flare affected a large volume, we found that the major dynamical processes occurred inside a part of the active region indicated by black boxes in Figure \ref{large fov}.

\begin{figure*}
\centering
  \includegraphics[width=1 \textwidth]{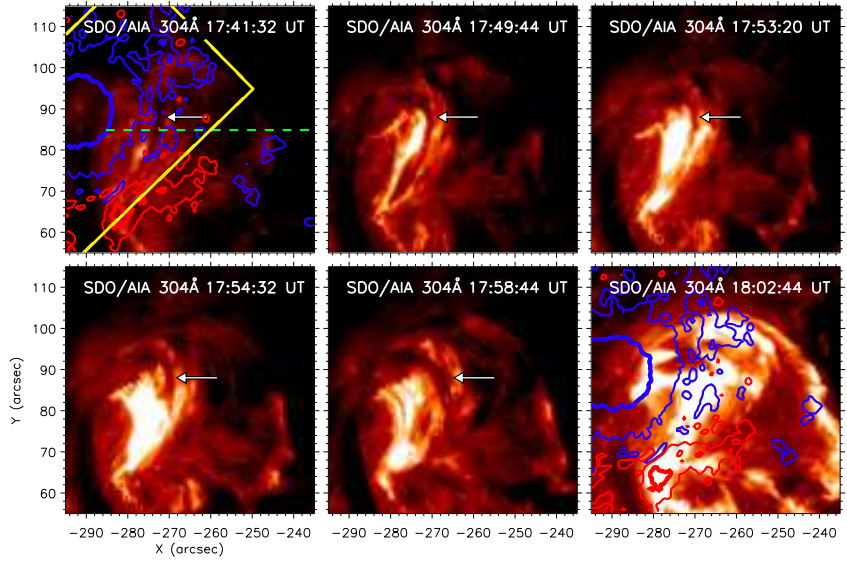}
      \caption{Time sequence of the SDO/AIA 304 \AA~images. The white arrows indicate the rising filament. The dashed green line indicates the slit for constructing the time-distance map (Figure \ref{time-distance}c). The yellow box corresponds to the FOV of the FISS.
              }
         \label{time sequence}
   \end{figure*}

Figure \ref{time sequence} shows the C5.6 main flare itself and its preflare seen in the SDO/AIA 304 \AA~images. The preflare started at 17:49 UT in-between the positive and negative poles of the moat region. Its brightness increased for about 5 minutes, then diminished. The impulsive phase of the main flare started at 18:01 UT. The 304 \AA~flare kernel was co-spatial with the positive and negative poles in the moat region and a part of the large negative polarity sunspot.
The most noteworthy feature seen in the 304 \AA~images is the filament indicated by the white arrows. This filament connects the negative and the positive poles of the moat region, and has an apparent length of 18 Mm and an apparent width of 1 Mm. It stayed stable until the preflare occurred near the middle of it. During the preflare activity, the filament slowly moved westward and accelerated, which likely reflects the projected component of the upward motion. After the main flare took place, the filament was no longer visible.

\section{Spectral Properties} 

\begin{figure*}
\centering
  \includegraphics[width=1 \textwidth]{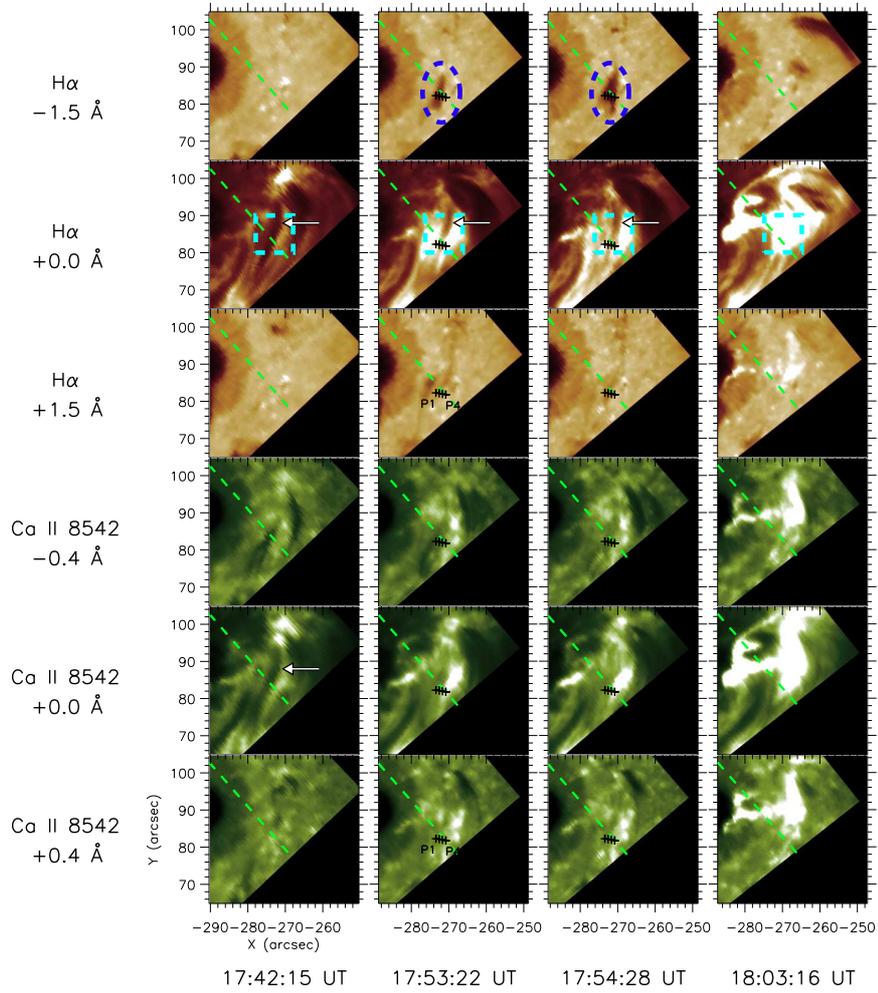}
      \caption{Time sequences of the FISS H$\alpha$ (brown) and Ca {\sc ii} 8542  \AA~(green) raster images at three wavelengths for each spectral line. In the top panel, the dashed blue ellipses mark the H$\alpha$ blueshifted broadening. The white arrows indicate the rising filament. The dashed cyan box is the region over which intensity is integrated to produce the light curve in Figure \ref{time-distance}a. The dashed green lines mark the slit position used for constructing the spectrograms displayed in Figure \ref{spectrogram}. The crosses indicate the P1 - P4 positions used for local spectra shown in Figure \ref{spectra}.
              }
         \label{time sequence Ha}
   \end{figure*}

The H$\alpha$ counterpart of the filament appears in the raster images constructed at several wavelengths from the FISS data as shown in Figure \ref{time sequence Ha}. It also appears as a filament clearly visible at the center of the H$\alpha$ and Ca {\sc ii} 8542 \AA~lines. Like the 304 \AA~filament, this H$\alpha$ filament was stable before the preflare, and began a dynamical change when the preflare started. The dynamic change is characterized by very large blueshift as can be inferred from the appearance of the elongated absorption feature at the far blue wing ($-1.5$ \AA) of the H$\alpha$ line (surrounded by the dashed blue ellipses in Figure \ref{time sequence Ha}). The feature, H$\alpha$ blueshifted broadening, had a width comparable to that of the rising filament and lasted for a few minutes. We note that the evolution of the filament seen in the Ca {\sc ii} 8452 \AA~line differs from that seen at the H$\alpha$ line. In the H$\alpha$ line center images, for instance, the filament is discernible throughout the preflare phase, while the filament soon vanishes in the Ca {\sc ii} 8542 \AA\ line images.

\begin{figure*}
\centering
  \includegraphics[width=1 \textwidth]{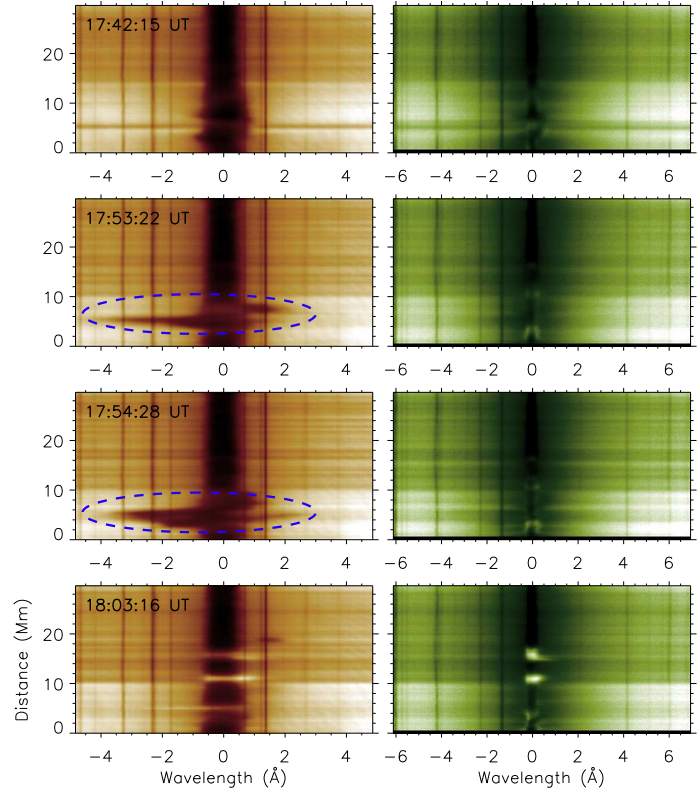}
      \caption{Examples of the FISS H$\alpha$ (left column) and Ca {\sc ii} 8542  \AA~(right column) spectrograms. The zero wavelength in the left (right) panels represents H$\alpha$ 6562.817 \AA\ (Ca {\sc ii} 8542.09 \AA). Each spectrogram was taken at the slit position marked by the green line in Figure \ref{time sequence Ha} at each time. The vertical axis indicates the distance from right lower end of the dashed green line in Figure \ref{time sequence Ha}. The dashed blue ellipses indicate the H$\alpha$ blueshifted broadening.
              }
         \label{spectrogram}
   \end{figure*}

The spectrograms of H$\alpha$ and  Ca {\sc ii} 8542 \AA~lines are shown in Figure \ref{spectrogram} where we mark the blueshifted broadening feature with dashed ovals. We note that this feature in the H$\alpha$ spectrogram looks somewhat similar to the rapid blue excursions (RBE) studied in detail by \citet{Voort2009} in that it appears in absorption at a wide range of wavelengths from the line center to the far blue wing. But our observed feature is physically different from the quiet Sun RBEs of \citet{Voort2009} in some aspects. Most of all, it extends much farther blueward, up to $-4.5$ \AA, than these well-known RBEs which typically extend blueward roughly up to $-2$ \AA.
We also note that the region of strong upward Doppler motion, the feature marked by the blue dashed ellipse in the top row of Figure 3, takes an elongated ellipse shape. The major axis of the ellipse must be along the filament axis, and the minor axis of the ellipse, comparable with the filament thickness. On the other hand, an RBE often exhibits a linear motion within a narrower width because it represents a mass flow along a spicule.
This also differs from the rapid red excursions (RREs), which again show jet-like features but in the downward motion  ({\it e.g.} \citealt{Sekse2013}).
Finally, we find that not only blue excursions, but also red excursions, up to 2 \AA, occur either nearby (at 17:53:22 UT) or at the same location (at 17:54:28 UT), composing a dynamically complex structure between 0 and 10 Mm in Figure \ref{spectrogram}.

Another important feature is that this broadening event is prominent only in the H$\alpha$ line and not clearly visible in the Ca {\sc ii} 8542 \AA~line. In the Ca {\sc ii} 8542 \AA~line, there is only a small change at the far blue wing. This means that the upwardly-moving H$\alpha$ material is almost transparent to the Ca {\sc ii} 8542 \AA~ line so that the observed Ca {\sc ii} 8542 \AA~ spectrum mostly comes from the unperturbed chromosphere below. This will happen when the upflow is heated to a temperature where the Ca {\sc ii} 8542 \AA~ line opacity no longer exists. There is no such clear-cut temperature, but here we give a rough estimate based on the literature. For instance, Figure 8.9 of \citet{Carroll1996} shows that the Ca {\sc ii} line forms at temperatures between 3000 K and $10^4$ K whereas H$\alpha$ forms in a much wider range from 5000 K to $5\times 10^4$ K. We thus suppose that the H$\alpha$ brightening region is heated to a temperature above $10^4$ K. Furthermore this plasma should co-exist with an even hotter plasma with temperature $\geq$ 5 $\times 10^{4}$ K to account for the He {\sc ii} 304 \AA\ brightening.

\begin{figure*}
\centering
  \includegraphics[width=1 \textwidth]{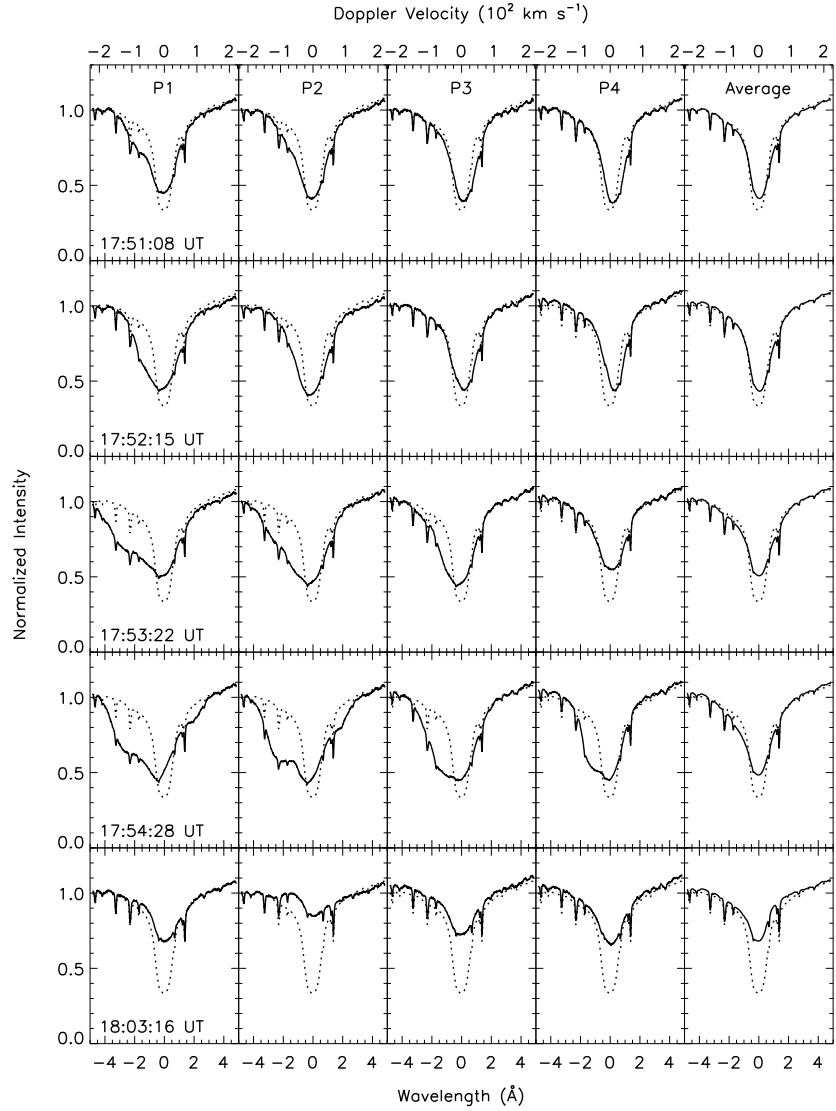}
      \caption{Temporal and spatial variation of the FISS H$\alpha$ spectra at the P1 - P4 positions marked in Figure \ref{time sequence Ha}.  The zero wavelength represents the H$\alpha$ line center at 6562.817 \AA. The dotted spectra are the average spectrum of the quieter region of the active region. The rightmost column shows the average line profiles integrated over the dashed cyan box in Figure \ref{time sequence Ha}.
              }
         \label{spectra}
   \end{figure*}

\begin{figure*}
\centering
  \includegraphics[width=1 \textwidth]{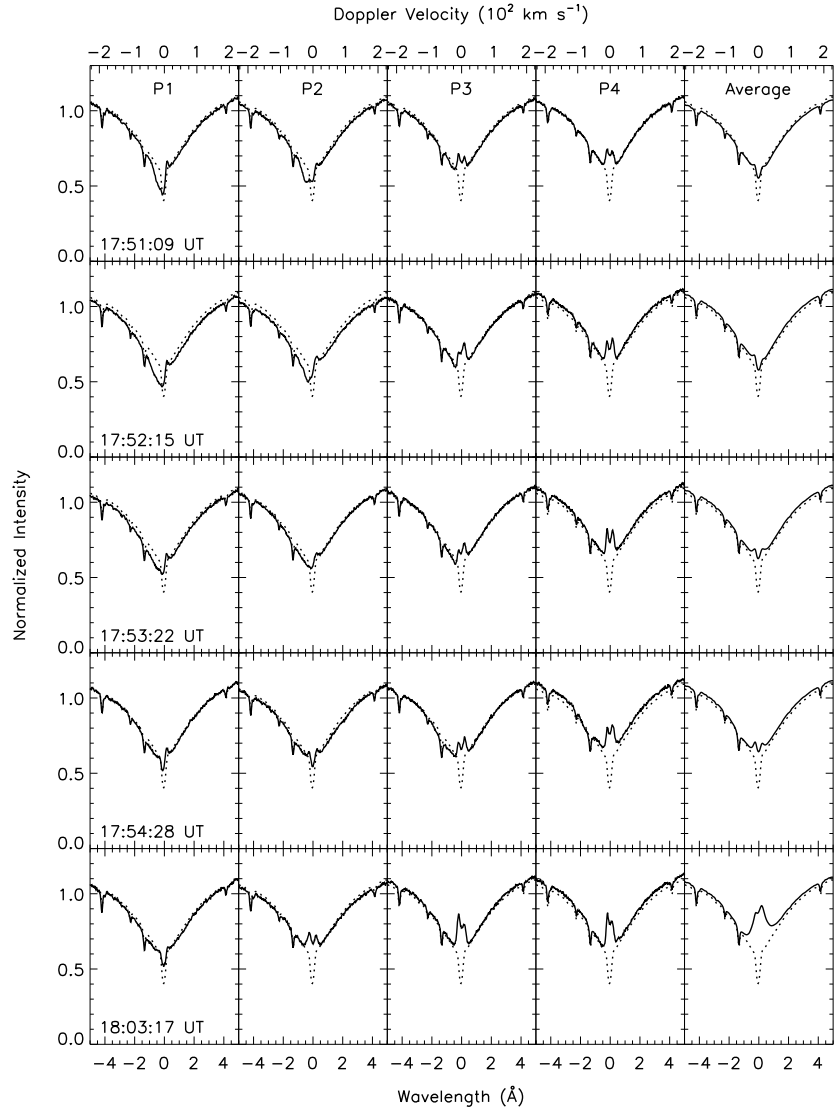}
      \caption{Temporal and spatial variation of the FISS Ca {\sc ii} 8542 \AA\ spectra. The panels follow the same format as Figure \ref{spectra}. The zero wavelength represents the Ca {\sc ii} 8542 \AA\ line center at 8542.09 \AA.
              }
         \label{spectra ca}
   \end{figure*}

Figure \ref{spectra} shows spatially resolved H$\alpha$ spectra at four time intervals from top to bottom panels. The first four columns show spectra determined at single locations, P1-P4. These locations are separated by 0.86 \arcsec, and each separation extends to 5.4 pixels. The rightmost column shows the average line profiles integrated over the box that covers the flaring area (the dashed cyan box in Figure \ref{time sequence Ha}). This strong blueshift broadening is confined to a spatially narrow region (P1-P2) over a short time interval ($\approx$ 6 min). If we look at the spatially averaged spectrum (right-most column) the strong blueshifted broadening could not have been recognized. This phenomenon is therefore hardly observable without high spatial and temporal resolution as well as spectral coverage. 
In Figure \ref{spectra ca} we plot the Ca {\sc ii} 8542 \AA\ line profiles in the same manner as Figure \ref{spectra}. Unlike the H$\alpha$ spectrum, the Ca {\sc ii} 8542 \AA\ line spectrum does not clearly show the blueshifted absorption feature. As mentioned we believe that this is due to the temperature being higher than $10^4$ K.

\begin{figure*}
\centering
 \includegraphics[width=1 \textwidth]{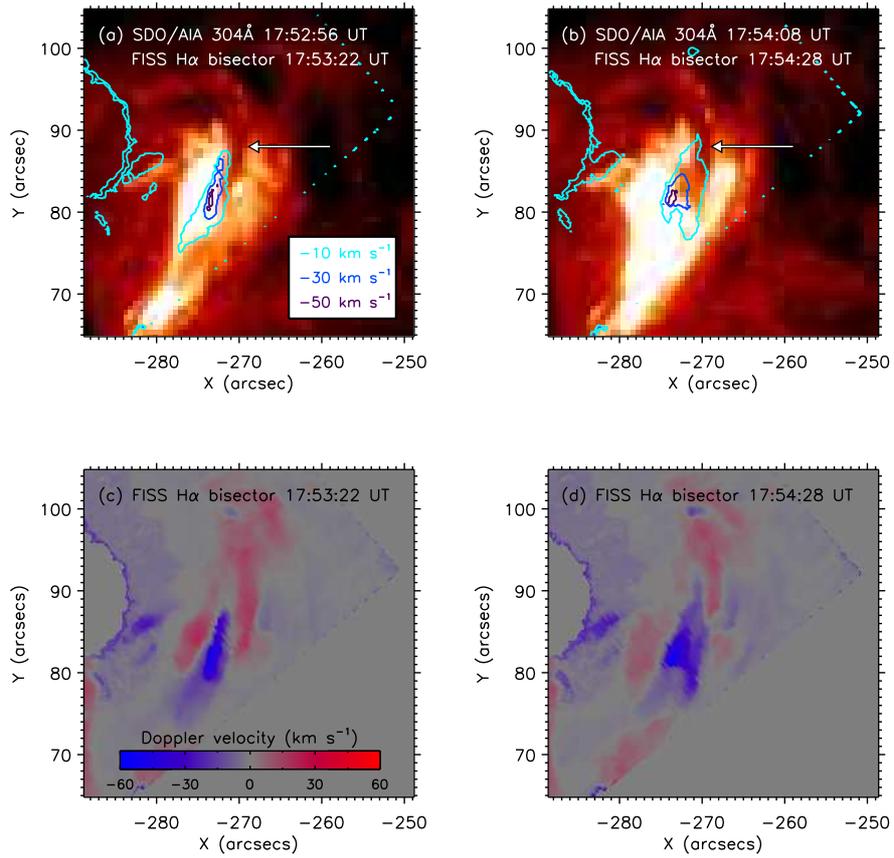}
      \caption{Spatial distributions of H$\alpha$ upward Doppler velocity. In the top panels the Doppler velocity maps (contours) are superposed on the SDO/AIA 304 \AA~images at two instants of the preflare phase. The white arrows indicate the rising filament. Bottom panels show the Doppler velocity maps only at the corresponding times.
              }
         \label{doppler velocity}
   \end{figure*}

In Figure \ref{doppler velocity}, we compare the H$\alpha$ Doppler velocity and the SDO/AIA 304 \AA~intensity map. The top two panels the show spatial distributions of H$\alpha$ Doppler velocity is plotted as contours on top of the SDO/AIA 304 \AA\ images at two instants of the preflare phase. The bottom panels show the Doppler maps as grayscale images at the corresponding times.
It turns out that the H$\alpha$ explosive event was co-spatial with the preflare
brightening and the rising filament.

Note that to determine the Doppler velocity in this figure, we used the bisector method \citep{Deubner1996, Chae2013a}. In this method, we typically select a level of the spectral line at which the bisector is taken as the measure of the Doppler shift. In the present case, we choose the 60 \% level of the quiet region continuum intensity, at which the blueshifted absorption feature of interest is well represented.
Because we used the bisector method, the Doppler velocities shown in the figure could be underestimated compared to those inferred from the blue ends of the excursions.

\section{Temporal Evolution}

\begin{figure*}
\centering
  \includegraphics[width=1 \textwidth]{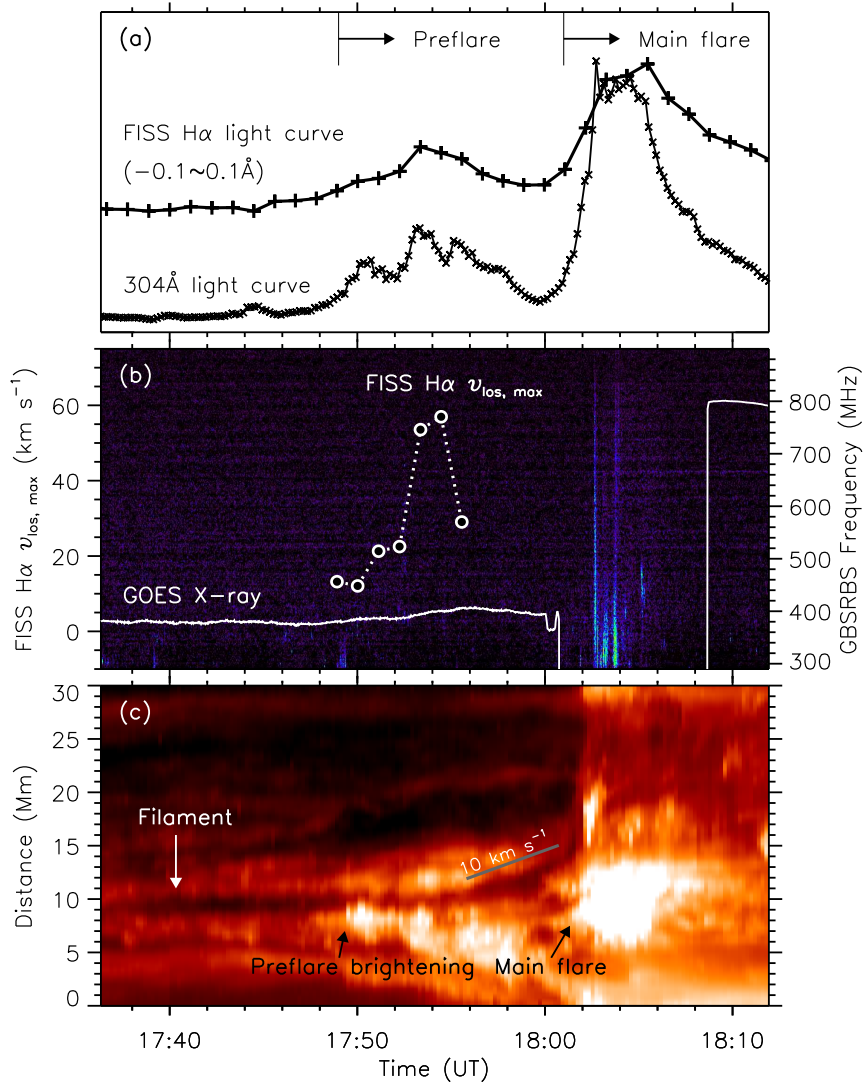}
      \caption{(a) Time profiles of  H$\alpha$ and SDO/AIA 304 \AA~measured inside the dashed sky blue box in Figure \ref{time sequence Ha}.
   (b) Time variations of the H$\alpha$ peak LOS velocity  (dotted line) and the GOES X-ray 1.0 - 8.0 \AA~flux (thin solid line) over-plotted on the radio dynamic spectrum from the GBSRBS (background image). The LOS velocities were determined by the bisector method applied to the FISS H$\alpha$ spectra in each time. The gap between 18:01 UT and 18:09 UT of the GOES X-ray is due to absence of data.
   (c) The time-distance map from SDO/AIA 304 \AA~image. The vertical axis indicates the distance from left end of the dashed green line in Figure \ref{time sequence}. The white arrow indicates the location of the filament.
              }
         \label{time-distance}
   \end{figure*}

Figure \ref{time-distance} presents the temporal variation of the H$\alpha$ blueshifted broadening in comparison with the development of the other features. The H$\alpha$ and SDO/AIA 304 \AA~lightcurves are represented in Figure \ref{time-distance}a. In Figure \ref{time-distance}b, the radio dynamic spectrum from the {\it Green Bank Solar Radio Burst Spectrometer} (GBSRBS) is shown as background image, and the maximum value of the H$\alpha$ Doppler velocity map at each time is plotted as symbols connected by the dotted line. Also shown as the white curve is the GOES soft X-ray lightcurve, which has a data gap during the maximum phase of the flare. Figure \ref{time-distance}c shows the time-distance map constructed from time series of the SDO/AIA 304 \AA~data along the slit defined in Figure 3 (the dashed green line).

The H$\alpha$ and SDO/AIA 304 \AA~lightcurves in \ref{time-distance}a clearly show that this preflare activity started about 6 min before the main phase of the flare.
The sizable LOS velocity of about $-10$ km s$^{-1}$ first appeared at 17:49 UT (\ref{time-distance}b), coinciding with the start of the preflare in Figure \ref{time-distance}a. The peak Doppler value increased to about $-60$ km s$^{-1}$ in about 6 minutes, which is during the period of the preflare activity of H$\alpha$ and 304 \AA. The H$\alpha$ blueshifted broadening lasted about 6 minutes during the preflare phase. Another noticeable property is that there is no enhancement in either soft X-rays or radio flux during the period of the H$\alpha$ blueshifted broadening.  Although hard X-ray data were unavailable for this event, the radio signals were detected in the main flare phase, which are of type III burst indicating high energy electrons escaping along open field lines. Therefore the data obtained in soft X-ray and radio wavelengths indicate this blueshift event is not associated with energetic electrons accelerated in the corona, although it is associated with plasma heating.

Figure \ref{time-distance}c shows the temporal variation of the transverse displacement of the 304 \AA~rising filament. This filament is clearly visible in absorption at about 17:40 UT (the white arrow in Figure \ref{time-distance}c). About 10 minutes later, it suddenly begins an oscillating motion. This instant corresponds to the time when the 304 \AA~preflare started and the H$\alpha$ blueshifted broadening first appeared. When the H$\alpha$ blueshifted broadening disappeared, the filament started to rise at noticeable transverse speeds of about 10 km s$^{-1}$. This slow-speed rise phase lasted about 6 minutes, and then the filament rapidly moved out of the field of view. The instant of transition from the slow to fast rise phase is very close to the start time of the main flare.

\section{Discussions} 

We focus on the nature and the role of the H$\alpha$ blueshifted broadening in the preflare stage, as is the main result of this study.

\subsection{Preflare Heating in the Chromosphere}

The strong H$\alpha$ Doppler broadening can be caused by either heating or nonthermal velocity, $\xi$, of the plasma or both. To determine these two quantities, we analyzed the H$\alpha$ spectrum of P2 at 17:54:28 UT, using the classical cloud model \citep{Beckers1964, Yang2013}. As a result we obtained the central wavelength of the absorption profile to be displaced by $-0.97$ \AA\ and the Doppler broadening of the line, $\Delta \lambda_{D} \approx 1.84$ \AA. The latter quantity is related to the temperature and nonthermal motion by $\Delta \lambda_{D}/\lambda = \sqrt{2k_{\rm B}T/m+\xi^2}/c$, where $k_B$ is the Boltzmann constant and $m$ is mass of the hydrogen atom. If we attribute this H$\alpha$ broadening entirely to thermal broadening ({\it i.e.}, $\xi=0$), the temperature should then be as high as $T \approx 4.3 \times 10^{5}$ K. At that temperature hydrogen atoms must be all ionized. It thus follows that the H$\alpha$ line broadening should be dominated by plasma motion. Of course, there must be plasma heating to some extent including the temperatures for H$\alpha$ and 304 \AA\ line formation. Since the He {\sc ii} 304 \AA\ line forms at $5\times 10^{4}$ K, the preflare brightening observed in the He {\sc ii} 304 \AA\ line indicates the presence of plasma at this temperature or higher. In order to explain the H$\alpha$ signature, a lower temperature plasma should co-exist, but the absence of a significant change in the Ca {\sc ii} 8542 \AA\ line requires the temperature to be above $10^4$ K. Another constraint is the absence of X-rays and microwaves during the preflare phase. This implies that the strong Doppler event is not accompanied by energetic electrons accelerated in the high corona, and the temperature should not exceed the typical coronal value. We also checked SDO/AIA data for a bright feature at 131 \AA\ but no significant counterpart at 94 \AA\ ($\log T\approx 6.8$) was found for the same region. We thus estimate that this preflare heating is mild and the plasma temperature is in the range $10^4$ K $ < T < 10^6$ K.

\subsection{Nature of the H$\alpha$  Blue-shifted Broadening}

If we adopt a reasonable temperature $T=10^{4}$ K in the cloud model, the observed H$\alpha$ broadening up to 1.84 \AA\ should indicate the presence of nonthermal motion of about $\pm$ 83 km s$^{-1}$. Since the displacement of the central wavelength of the absorption profile is $-0.97$ \AA\ or equivalently, $-45$ km s$^{-1}$, the whole motion lies in a wide velocity range from $-$130 km s$^{-1}$ to 38 km s$^{-1}$. These values are comparable to the highest speeds of other similar chromospheric phenomena as a preflare activity, for instance, blue-shifted velocity of $-140$ km s$^{-1}$ \citep{Canfield1990}, limb observation of flow motions with speed of 100 km s$^{-1}$ \citep{Sterling2011}, and a filament speed of $\approx$ 600 km s$^{-1}$ from the Si {\sc iv} line in the transition region \citep{Kleint2015}.


We interpret this strong H$\alpha$ broadening as an explosive H$\alpha$ event. The bulk motion itself is not so strong as the core of the line profile shifts only by $\approx$ 0.5 \AA ~(Figure \ref{spectra}). The broadening is, however, strong and implies the presence of velocities in the wide range from $-$130 to 38 km s$^{-1}$, implying an explosive event resulting in random motion with a large velocity dispersion. An alternative mechanism to have an H$\alpha$ broadening without any explosion is a chromospheric condensation in which the steep velocity gradients in the flaring chromosphere can modify the wavelength of the central reversal in the H$\alpha$ line profile, and the blue asymmetry could be generated as the maximum opacity shifts to shorter wavelengths ({\it e.g.}, \citealt{Gan1993}, \citealt{Heinzel1994}). We however note that the blue asymmetry in the latter model is due to the shift of an emission peak to shorter wavelengths, whereas in this event are absorption feature shifts to shorter wavelengths.

In this interpretation, we should mention that the velocities determined from the bisector method carry an uncertainty for a flaring atmosphere. The bisector attributes the changes in the absorption line profiles intensities to the line width and LOS velocities only, while intensity changes of the flaring atmosphere are highly dependent on the density, opacity, and the source function. For instance, a recent radiative hydrodynamic code (RADYN) simulation of flare H$\alpha$ line shows that an asymmetric line profile with a deformed blue wing is not necessarily associated with plasma upflow \citep{Kuridze2015}.  The same uncertainty may apply to the present event, although it is a much weaker preflare activity as compared with the M1.1 flare targeted by the RADYN simulation. It is important that the line asymmetry significantly changes with time and space as shown in Figure \ref{spectra}. The characteristic scale of such spectral variation is about the width of the filament, and we presume that this explosive event occurred within the filament.

Note also that this blueshifted event differs from the RBE, small-scale jets observed on the disk in the quiet Sun \citep{Voort2009}. Their typical width is around 250 km and the length is around 4 Mm. Their Doppler velocities are estimated as $\approx$ 30--40 km s$^{-1}$. The present blueshifted event is comparable to RBE in time but spectrally much wider and occurs in the pre-flare stage. Since this H$\alpha$ blueshifted broadening occurred adjacent to the regions of the EUV brightening, it is possible that that the local pressure was impulsively enhanced in the 304 \AA\ brightening region, resulting in the explosive motion of  H$\alpha$ material with a large velocity dispersion within the hot region.

\subsection{Relation to Other Flare Signatures}

Apparently the blueshifted H$\alpha$ broadening is temporally and spatially coincident with the start of the slow rise motion of a filament. It is, however, common in flare observations that these two phenomena occur together, and we are unsure  which one triggers the other. If any increases in the X-ray or microwave fluxes were found at $\approx$ 17:50 UT, we may have concluded that both of them were triggered by the energy deposition into the chromosphere and subsequent heating by energetic electrons accelerated in the corona. As mentioned in section \S 2, there was no X-ray counterpart detected by the RHESSI for this event. What is certain is that the slow rise motion ($\approx$ 10 km s$^{-1}$) of the filament started after the H$\alpha$ blueshifted broadening, and it led to the main flare (Figure \ref{time-distance}c). In this sense we compare this event to the two stage eruption pattern for flares \citep{Wang1993} in which the preflare brightening represents the first magnetic reconnection in the lower atmosphere, resulting in the H$\alpha$  blueshifted broadening, and the second reconnection responsible for the main phase of the flare occurs in the corona. We however cannot provide evidence for a causal relationship between the two stages with the current data alone.

\section{Summary}

We have analyzed H$\alpha$ blueshifted broadening associated with the preflare activity of the solar flare SOL2011-09-29T18:08, exploiting the H$\alpha$ spectrographic observations with the FISS along with the corresponding SDO/AIA EUV images. We find unusually strong H$\alpha$ blueshifted broadening in the preflare stage using the H$\alpha$ line profile measured at 512 spectral positions with spectral sampling of 0.019 \AA~ on maps of spatial intervals of 0.16 \arcsec\ per pixel.
Our results are summarized as follows:

\begin{enumerate}

\item
The foremost preflare activity was the brightening at 304 \AA\ around a filament that started from 17:49 UT. H$\alpha$ brightness distribution resembles that of 304 \AA. There was no counterpart in X-rays and radio bursts, and no significant change in Ca {\sc ii} line spectrum. From these, we estimate the temperature range of this preflare heating as $10^4$ K $ <T < 10^6$ K.

\item
The preflare H$\alpha$ brightening is simultaneous and co-spatial with the 304 \AA\ brightening. The spectral shape of the H$\alpha$ broadening indicates a mixture of random velocities in a wide range between $-130$ km s$^{-1}$ and 38 km s$^{-1}$ at the maximum time 17:54 UT.

\item
 The strong H$\alpha$ upflow lasted about 6 minutes, while the EUV brightening continued in the low chromosphere. It was at the maximum of the H$\alpha$ blueshifted broadening when the filament started to rise slowly ($\approx$ 10 km s$^{-1}$). Because of the temporal and spatial coincidence, we interpret this event as the impulsive increase of local pressure in the 304 \AA\ brightening region resulting in the explosive motion of  H$\alpha$ materials with a large velocity dispersion within the hot region.

\item
The filament rises from the low chromosphere at a low speed of $\approx$ 10 km s$^{-1}$, until it reaches a certain coronal height to further accelerate at the start of the main phase of the flare (18:01 UT). This pattern is a familiar feature of the two-step eruption model, where the slow rise followed by fast eruption occurs under a rapid change of magnetic field in the corona.

\end{enumerate}

In this interpretation, we expect that FISS observations of preflare H$\alpha$ blueshifted broadening will continue to be important for finding clues to the early activation of solar flares.

\begin{acks}
We thank the NASA/SDO team for the photospheric magnetograms and UV/EUV filtergrams, and NASA/Fermi team for the GBM data. We thank Dr. Stephen White for the GBSRBS radio data. This work was supported by the National Research Foundation of Korea (NRF-2012R1A2A1A03670387). HW is supported by US NSF under grants AGS 1348513 and 1408703, and NASA under grant NNX13AG13G. E.-K, Lim is supported by the Planetary system research for space exploration from KASI. JL is supported by the BK21 Plus Program (21A20131111123) funded by the Ministry of Education (MOE, Korea) and National Research Foundation of Korea (NRF).
\end{acks}



\bibliographystyle{spr-mp-sola}
\bibliography{ms4}


%
%

%
%

%
%

%
%

%
%
%

%
%
%
%

\end{article}

\end{document}